\documentstyle[referee]{mn}
\def\sc33{OGLE-1999-BUL-32}
\newcommand{\reference}{\bibitem}
\def\mnras{MNRAS}
\def\araa{ARAA}
\def\aap{A\&A}
\def\apj{ApJ}
\def\plotone#1{\centering \leavevmode
\epsfxsize=\columnwidth \epsfbox{#1}}

\def\thetaE{\theta_{\rm E}}
\def\Amax{A_{\rm max}}
\def\mas{\,{\rm mas}}
\def\vt{\tilde{v_t}}
\def\beq{\begin{equation}}
\def\eeq{\end{equation}}
\def\bey{\begin{eqnarray}}
\def\eey{\end{eqnarray}}
\def\kms{{\rm \,km\,s^{-1}}}

\def\kpc{\,{\rm {kpc}}}

\def\chisq{\chi^2}

\def\tE{t_{\rm E}}
\def\rEt{\tilde{r}_{\rm E}}

\def\u0{u_0}

\def\Ds{D_{\rm s}}
\def\Dl{D_{\rm l}}
\def\mI0{m_{I,0}}

\def\fl{f_{\rm L}}

\def\pirel{\pi_{\rm rel}}

\newcommand{\up}[1]{\ifmmode^{\rm #1}\else$^{\rm #1}$\fi}

\newcommand{\arcd}{\ifmmode^{\circ}\else$^{\circ}$\fi}
\newcommand{\arcm}{\ifmmode{'}\else$'$\fi}
\newcommand{\arcs}{\ifmmode{''}\else$''$\fi}

\input epsf
\begin{document}

\title[
Optical Gravitational Lensing Experiment. Evidence for a Stellar
Black Hole?
]
{
Optical Gravitational Lensing Experiment.
\sc33: the Longest Ever Microlensing Event -- Evidence for a
Stellar Mass Black Hole?
}
\author[Mao, Smith, Wo\'zniak, Udalski et al.]
{
Shude Mao$^{1,2}$,
Martin C. Smith$^1$,
P. Wo\'zniak$^{3}$,
A. Udalski$^4$, M. Szyma\'nski$^4$,
\newauthor M. Kubiak$^4$, G. Pietrzy\'nski$^{5,4}$,
I. Soszy\'nski$^4$, K. \.Zebru\'n$^4$
\thanks{e-mail: (smao,msmith)@jb.man.ac.uk, wozniak@lanl.gov,
(udalski,msz,mk,pietrzyn,soszynsk,zebrun)@astrouw.edu.pl
}
\thanks{
Based on observations obtained with the 1.3 m Warsaw
Telescope at the Las Campanas Observatory of the Carnegie
Institution of Washington.
}
\\
\smallskip
$^{1}$ Univ. of Manchester, Jodrell Bank Observatory, Macclesfield,
Cheshire SK11 9DL, UK \\ 
$^{2}$ Princeton University Observatory, Princeton, NJ 08544-1001, USA \\
$^{3}$ Los Alamos National Laboratory, MS D436, Los Alamos,
NM 87545, USA \\
$^{4}$ Warsaw University Observatory, Al. Ujazdowskie 4,
00-478 Warszawa, Poland \\
$^{5}$ Universidad de Concepci{\'o}n, Departamento de Fisica,
Casilla 160--C, Concepci{\'o}n, Chile
}
\date{Accepted ........
      Received .......;
      in original form ......}

\pubyear{2001}

\maketitle
\begin{abstract}
We describe the discovery of the longest microlensing event ever
observed, \sc33, also independently identified by
the MACHO collaboration as MACHO-99-BLG-22. This unique event
has an Einstein radius crossing time of 641 days. The high
quality data obtained with difference image analysis
shows a small but significant parallax signature. This parallax effect
allows one to determine the Einstein radius projected onto the observer
plane as $\rEt\approx 29.2$AU. The transverse velocity projected onto the
observer plane is about $79\kms$. We argue that the lens is likely to be 
have a mass of at least a few solar masses, i.e., it could be a 
stellar black hole. The black hole hypothesis can be tested using the
astrometric microlensing signature 
with the soon-to-be installed Advanced 
Camera for Surveys on board the Hubble Space Telescope. Deep X-ray
and radio images may also be useful for revealing the nature of the object.
\end{abstract}


\section{Introduction}

Gravitational microlensing is rapidly becoming an important
astrophysical tool (for a review, see Paczy\'nski 1996). The unique strength
of this technique is that it provides a mass-selected
sample for a variety of astrophysical applications, such as studying the
Galactic structure and mass functions in the local group.
So far, over one thousand
microlensing events, mostly toward the Galactic bulge, have been
discovered (e.g. Alcock et al. 2000; Wo\'zniak et al. 2001; Bond et
al. 2001). Most ($\sim 90$ per cent)
microlensing events are well described by the standard shape
(e.g. Paczy\'nski 1986). Unfortunately, from these light curves, one
can only derive a single physical constraint, namely the Einstein radius
crossing time, which involves the lens mass, various distance measures
and relative velocity (see \S 3). This degeneracy means that the lens
properties cannot be uniquely inferred, thus making the
interpretation of the microlensing results ambiguous.

 The parallax
microlensing events are one class of exotic microlensing events 
that allow this degeneracy to be partially removed. The parallax effect
we discuss here arises when the
event lasts long enough that the Earth's motion can no longer be
approximated as rectilinear during the event (Gould 1992;
see also Refsdal 1966 for a related effect).
Unlike the light curves for the standard events which are symmetric,
these parallax events often exhibit asymmetries in their light curves due 
to the  motion of the Earth around the Sun. These events
allow one to derive the physical dimension of the
Einstein radius projected onto the observer plane
and hence the lens degeneracy can be partially lifted.

A number of parallax microlensing events have been reported
in the literature (Alcock et al. 1995; Mao 1999; 
Soszy\'nski et al. 2001; Bond et al. 2001; see also 
Bennett et al.\ 1997).  Smith, Mao \& Wo\'zniak (2001) recently 
developed a method to systematically search for
parallax signatures in the OGLE-II microlensing candidates
found by Wo\'zniak et al. (2001).
We have uncovered several parallax candidates 
in this database. One of these, \sc33, turns out to 
be the longest microlensing event ever observed.
The purpose of this paper is to analyze
this unique event in some detail. We argue that this event is
likely to be caused by a stellar mass black hole; other black hole candidates
from microlensing have been reported in conference
abstracts (Bennett et al. 1999; Quinn et al. 1999). The
outline of the paper is as follows. In \S2 we briefly describe
the observations, data reduction and our parallax search algorithm,
in \S3 we describe our model for this spectacular microlensing
event, and in \S4 we propose future observations that can further test
our model, particularly with the Hubble Space Telescope (HST).

\section{Observations, Data Reduction and Selection Procedure}

All observations presented in this paper were carried out during the
second phase of the OGLE experiment with the 1.3 m Warsaw telescope
at the Las Campanas Observatory, Chile. The observatory is operated
by the Carnegie Institution of Washington. The telescope was
equipped with the `first generation' camera with a SITe 3
2048$\times$2048 pixel CCD detector working in the drift-scan mode.
The pixel size was 24$\mu$m, giving the scale of 0.417\arcs per pixel.
Observations of the Galactic bulge fields were performed in the
`medium' speed reading mode with the gain 7.1 e$^{-}$ ADU$^{-1}$
and readout noise about 6.3 e$^{-}$. Details of the instrumentation
setup can be found in Udalski, Kubiak \& Szyma\'nski (1997).
The majority of the OGLE-II frames were taken in the {\it I}-band,
roughly 200-300 frames per field during observing seasons 1997--1999.
Udalski et al. (2000) gives full details of the standard OGLE
observing techniques, and the DoPhot photometry
(Schechter, Mateo \& Saha 1993) is available from the OGLE
web site at {\it http://www.astrouw.edu.pl/\~\,ogle/ogle2/ews/ews.html}.

Wo\'zniak et al. (2001) searched for microlensing events 
in the three year OGLE-II bulge data analyzed using difference image analysis.
The difference image analysis 
pipeline is designed and tuned for the OGLE bulge data
(Wo\'zniak 2001), and is based on the algorithm from Alard \& Lupton (1998)
and Alard (2000).
The difference image analysis pipeline returned a catalog of over 200,000
candidate variable objects, from which 520 microlensing candidates were
identified using a combination of an algorithmic search,
visual inspections and a cross-correlation with the candidates
identified by Udalski et al. (2000) from the DoPhot analysis.
The details can be found in
Wo\'zniak et al. (2001) and will not be repeated here.

We then searched for parallax microlensing events using the method
developed in Smith et al. (2001). Here we outline the
prescription. In the first step, we fit each microlensing light curve
with both the standard model and the parallax model (see \S3 for
the procedure applied to \sc33). The
events that show significant improvements with the incorporation
of the parallax effect are then recorded and subjected to further
studies. Among the recorded events, we then select those events
for which the peak is at least 30 times higher than the noise level and
the time interval during which the microlensing variability is at least
$3\sigma$ above the noise level is longer than 100 days.
These two filters properly account for the fact that
(subtle) parallax signatures are most likely to be detectable in
long-duration events and those with high signal-to-noise ratio. We found
this prescription to be successful. Several
good candidates and a number of marginal ones were identified.
We refer the readers to Smith et al. (2001) for further details.

\begin{figure}
\plotone{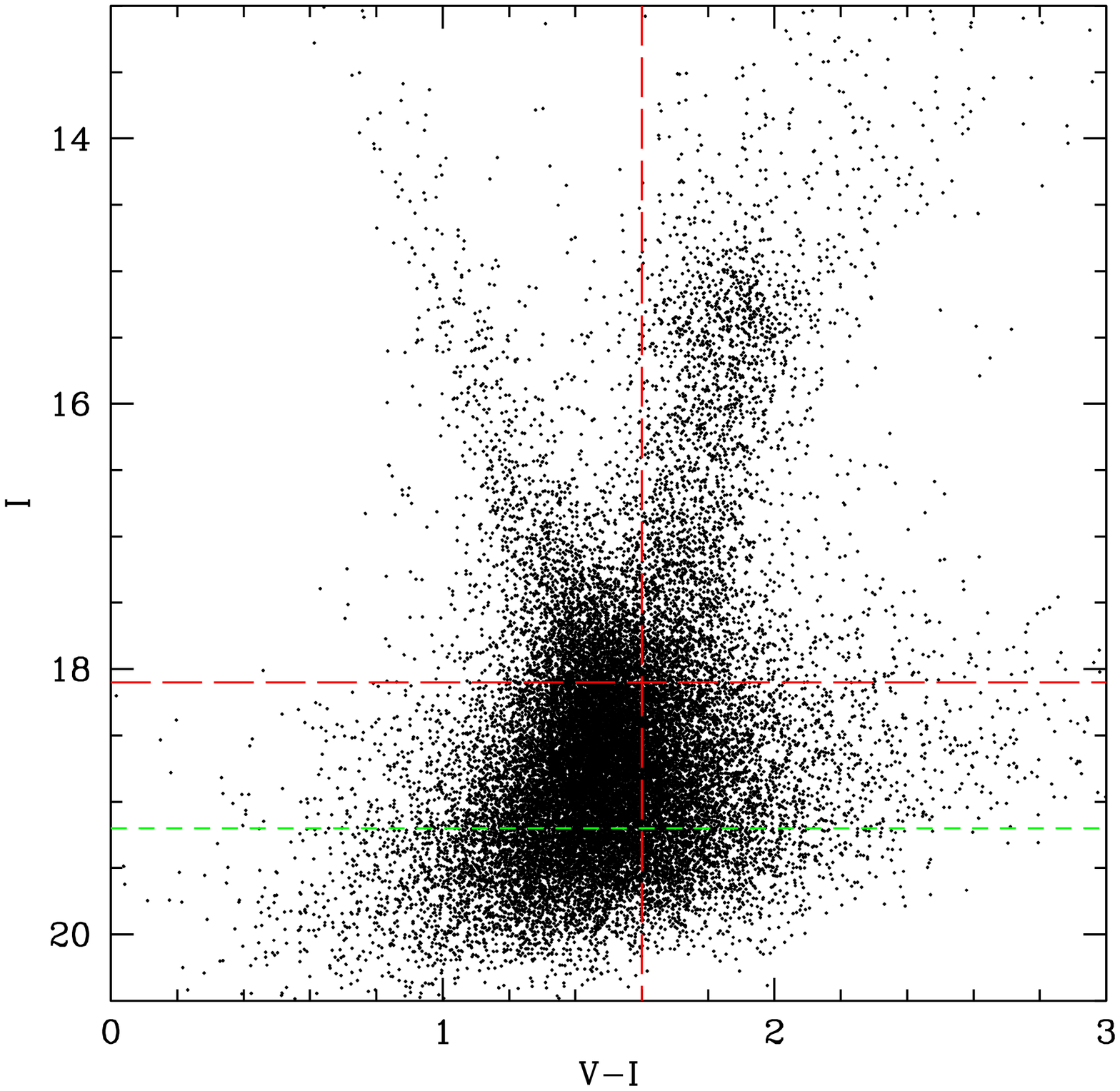}
\caption{The colour-magnitude diagram for the $3.5\arcmin \times
8\arcmin$ stellar field around \sc33. The short-dashed line
indicates the {\it I}-band baseline magnitude for the lensed star, while
the two long-dashed lines indicate
the magnitude and the colour of the total light from the lensed star
plus nearby blend(s).
\label{fig:cmd}
}
\end{figure}

\begin{figure}
\plotone{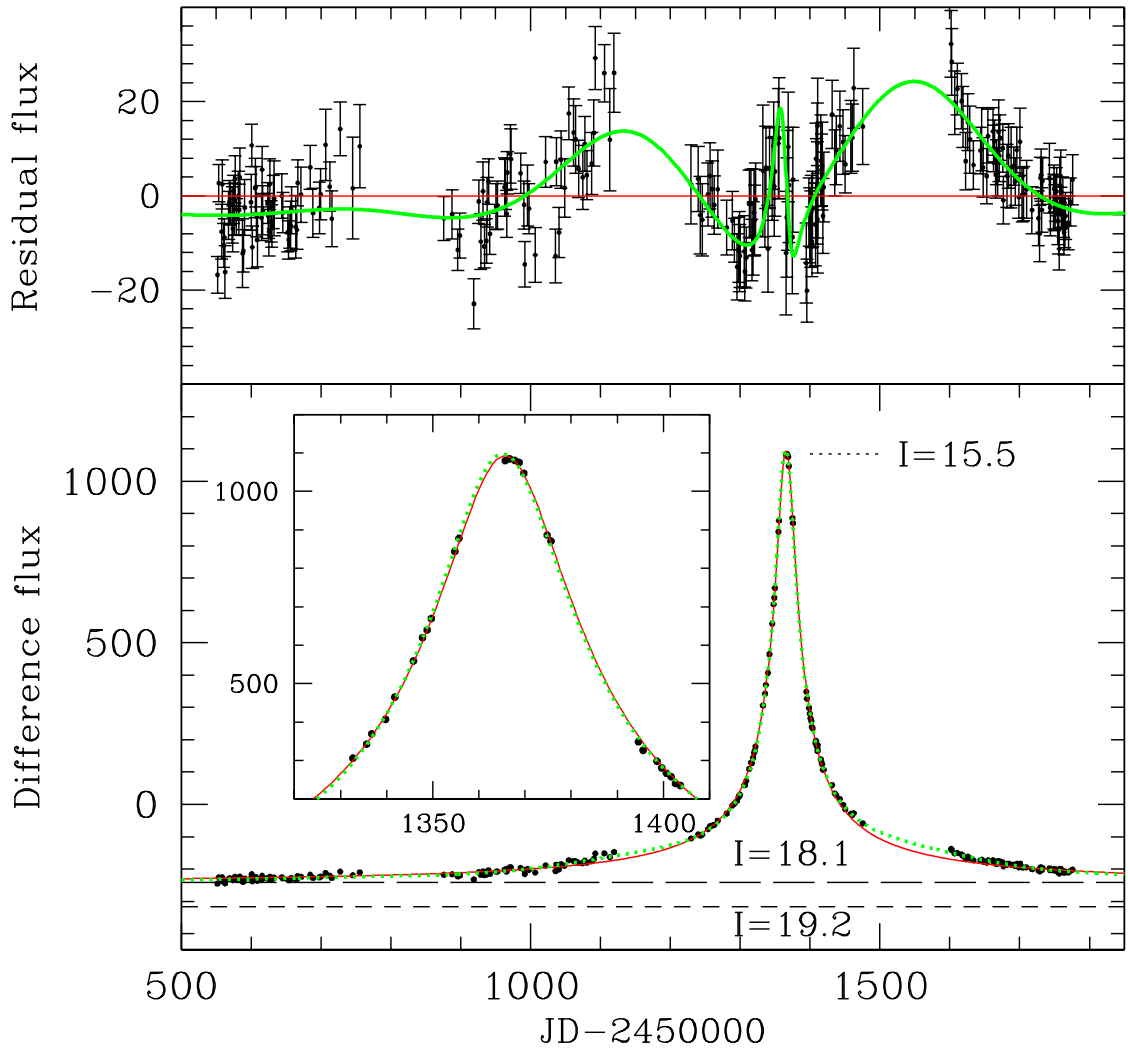}
\caption{The {\it I}-band light curve for \sc33 from difference image
analysis. The solid and dotted lines are for the standard and parallax fits,
respectively. The short-dashed line shows the baseline flux of the lensed
star while the long-dashed line shows the total baseline flux
of the lensed star and nearby blend(s).
The approximate {\it I}-band magnitudes are indicated for these two
baselines, together with the peak {\it I}-band magnitude.
The inset shows the the light curve close to the peak.
The top panel shows the residual flux (the observed data points
subtracted by the standard model). Clearly the standard model shows
systematic discrepancies. The curved solid line shows the
prediction of the parallax model.
\label{fig:lc}
}
\end{figure}

In this algorithmic search, one microlensing event in Wo\'zniak et al.'s
catalog, sc33\_3764, passed all our criteria. The microlensing variability
was in fact first identified
in real-time by the MACHO alert system as MACHO-99-BLG-22; it was also
detected by the OGLE early-warning-system as
OGLE-1999-BUL-32. The star, however, first escaped detection as a microlensing
candidate (or even as a `transient',
see Wo\'zniak et al. 2001) in the difference image analysis, because 
the star never
reached a `constant' baseline (see Fig. \ref{fig:lc}). The event was
recovered by cross-correlating the variable stars with the microlensing
candidates found by Udalski et al. (2000).
Throughout this paper, we shall refer to this event as
\sc33, following the notation of Udalski et al. (2000). The position
of the star is RA=18:05:05.35, and DEC=$-$28:34:42.5 (J2000). The
Galactic coordinates are $l=2^\circ.460, b=-3^\circ.505$. The
DoPhot photometry and finding chart for the star are available online.
\footnote{http://www.astrouw.edu.pl/\~\,ogle/ogle2/ews/1999/bul-32.html;
\\
ftp://darkstar.astro.washington.edu/macho/Alert/99-BLG-22/}
The total {\it I}-band magnitude of the
lensed star and the nearby blend(s) is about $I\approx 18.1$
(uncertain by about 0.05mag, Wo\'zniak et al. 2001). The baseline
magnitude of the lensed star alone is about $I \approx 19.2$ 
(see \S3).
There are several $V$-band frames in the OGLE-II database when the composite
was fainter than $I=16.6$ magnitude. The average $V-I$ colour of the 
composite is about 1.6.
Fig. \ref{fig:cmd} shows the colour-magnitude diagram for the stars
within a field of view $3.8\arcmin \times 8\arcmin$ around \sc33.  From
this figure, it is clear that the magnitude
and colour of the total light is similar to most stars in this
direction. This is also true for the magnitude of the lensed star,
although its colour is unknown. Therefore \sc33 is entirely consistent
with being approximately at the Galactic centre. In the same
diagram, the red clump stars around $I=15.3$ and $V-I=1.8$ are 
clearly visible.

The (online) DoPhot photometry is quite noisy, because the lensed star 
is heavily blended with nearby star(s) (see \S3),
the fluctuations in the seeing make it difficult for 
DoPhot to disentangle the relative contributions.
In fact, it is so noisy that the time-scale
of this event is hard to determine with the DoPhot photometry. In 
contrast, the difference image analysis automatically subtracts out any
blending. As a result, the errors are much reduced
and the number of usable images is also increased. Both improvements
are crucial for determining the long-duration nature
of the event and, more importantly,
for detecting the subtle parallax effect. Initially we analysed just the
three season data from 1997 to 1999 available online (Wo\'zniak et
al. 2001). However, the parallax model predicts deviations from the 
standard model in the 2000 season and to test this we subsequently 
analysed the data from this season. Reassuringly, this confirmed the 
prediction of our parallax model. The four-season
data from the difference image analysis
is shown in Fig. \ref{fig:lc}\footnote{The data are available
at http://www.jb.man.ac.uk/\~\,smao/bul33.dat}. In total, there are 246 data points in the light
curve. In the next section, we present both the best standard
and parallax model for this unique event.

\section{Model}

We first fit \sc33 with the standard single microlens model.
In this model, the (point) source, the lens and the observer are all
assumed to move with constant spatial velocities. The standard light curve,
$A(t)$ is given by (e.g. Paczy\'nski 1986):
\beq \label{amp}
A(t) = {u^2+2 \over u \sqrt{u^2+4}},~~
u(t) \equiv \sqrt{\u0^2 + \tau(t)^2},
\eeq
where $\u0$ is the impact parameter (in units of the Einstein radius) and 
\beq \label{tau}
\tau(t) = {t-t_0 \over \tE}, ~~ \tE = {\tilde{r}_{\rm E} \over \tilde{v}},
\eeq
with $t_0$ being the time of the closest approach (maximum
magnification), $\tilde{r}_{\rm E}$ the Einstein radius projected onto the 
observer plane, $\tilde{v}$ the lens transverse velocity relative to 
the observer-source line of sight, also projected onto the observer
plane, and $\tE$ the Einstein radius crossing time.
The Einstein radius projected onto the observer plane is given by
\beq \label{rE}
\tilde{r}_{\rm E} = \sqrt{4 G M \Ds x \over {c^2 (1-x)}},
\eeq
where $M$ is the lens mass, $\Ds$ the distance to the source and
$x=\Dl/\Ds$ is the ratio of the distance to the lens and the distance
to the source. Eqs. (\ref{amp}-\ref{rE}) shows the well-known
lens degeneracy, i.e., from a measured $\tE$,
one can not infer $\tilde{v}$, $M$ and $x$ uniquely even if the source distance
is known. 

The flux difference obtained from difference image analysis can be
written as
\beq \label{eq:f}
f(t) = \fl \left[A(t) - 1\right] +\Delta f,
\eeq
where $\fl$ is the baseline flux of the lensed star, and $\Delta f
\equiv f_0-f_R$ is the difference between the baseline flux ($f_0$)
and the flux of the reference image ($f_R$). $f_0$
includes the (unmagnified) 
flux of the lensed star and blended star(s), if present.
Note that in general $\Delta f$ does not have to be zero or even
positive as the reference image can be brighter than the 
true baseline image ($f_R>f_0$). For \sc33, the reference
image flux is $f_R=359.5$ (Wo\'zniak et al. 2001). Therefore, to fit
the {\it I}-band data with the standard model, we need five
parameters, namely, $\fl$, $\Delta f$ (or $f_0$), $\u0, t_0$, and $\tE$.
Best-fit parameters (and their errors)
are found by minimizing the usual $\chisq$ using the MINUIT program in the
CERN library$\footnote{http://wwwinfo.cern.ch/asd/cernlib/}$.

Our attempts to fit the light curve with the standard model reveal an 
ambiguity. This is due to the degeneracy between $\fl$, $\u0$ and $\tE$ 
for a heavily blended light curve (Wo\'zniak \& Paczy\'nski 1997). 
In such cases,  only the combinations $\u0\tE$ and $\fl/\u0$ are well
determined, but not $\u0, \tE$ and $\fl$ individually. 
If the parameter $\u0$ is left unconstrained for this event, then a $\u0$
value close to zero is formally preferred, with $\chi^2=524.4$ 
for 241 degrees of freedom.
However, such a perfect alignment is statistically unlikely. 
For illustrative purposes, in Fig. \ref{fig:lc}, we show the best fit with
$\u0$ fixed to be 0.01, which has a slightly worse $\chi^2=530.7$
than the best fit with $\u0$ left unconstrained. The fit parameters are
presented in Table 1. 
The top panel in Fig. \ref{fig:lc} shows the difference between 
the data points and the standard model.
Clearly the observed light curve shows systematic 
deviations from the model. Quantitatively,
the $\chi^2$ per degree of freedom is about $\approx 2.2$,
which is unacceptably large. Since the microlensing variability
can be clearly seen
over at least four years, during which time the Earth has moved through
four orbits, it is natural to ask whether the incorporation
of the parallax effect will remove the inconsistency. We show next that
this is indeed the case.

To account for the parallax effect, we follow the natural formalism of
Gould (2000)
and describe the lens trajectory in the ecliptic plane. This requires 
two further parameters, namely the projected Einstein radius onto 
the observer plane, and an angle $\psi$ in the ecliptic plane,
which is defined as the angle between
the heliocentric ecliptic $x$-axis and the normal to the
trajectory
(This geometry is illustrated in Fig.~5 of Soszy\'nski et al. 2001).
Once these two parameters are specified, the resulting lens trajectory in the 
ecliptic plane completely determines the separation between the lens and the 
observer (i.e., the quantity which is analogous to the standard model's $u_0$ 
parameter from eq. 1). This allows the light curve to be
calculated; the complete prescription is given in Soszy\'nski et
al. (2001), to which we refer the reader for further technical 
details (see also Alcock et al. 1995; Dominik 1998).
For the parameters $\fl$, $\Delta f$, $\u0, t_0$, and $\tE$, we take the fit
parameters from the standard model as the initial guesses, while $\rEt$
and $\psi$ are arbitrarily chosen for a number of combinations to search
for any degeneracy in the parameter space.
The best-fitting parameters are again found by minimizing the $\chi^2$. 
Notice that in the parallax
model, $\u0$ and $t_0$ describe the closest approach and the
corresponding time
of the lens trajectory with respect to the Sun in the ecliptic plane. 
They no longer have straightforward intuitive interpretations
as analogous parameters in the standard model, due to geometric projections
and the parallax effect. For example, the closest approach in the
ecliptic plane is in general not the closest approach in the lens plane,
and hence does not correspond to the peak of the light curve.

\begin{table}
\begin{center}
\caption{The best standard model (first row) with the impact parameter
$\u0$ fixed to be 0.01 and the best parallax model (second row) for \sc33.
The parameters are explained in \S3.
} 
\vspace{0.3cm}
\begin{tabular}{ccccccccc}
Model & $t_0$ & $\tE$ (day) & $\u0$ & $\fl$
& $\Delta f$ & $\psi$ & $\rEt$ (AU) & $\chisq$/dof \\
\hline
$S$ & $ 1365.7\pm       0.08$ &
$    1495.9\pm     8.7$ &
$     0.01       $ &
$    13.48\pm 0.05$ &
$   -242.5\pm       0.48$ &
--- & --- & 550.62/241
\\
P & $ 1322_{-55}^{+18}$ &
$    641_{-55}^{+70}$ &
$     0.08_{-0.03}^{+0.24}$ &
$    43.3 \pm 5.1$ &
$   -240.9_{-1.6}^{+1.4}$ &
$    3.38_{-0.1}^{+1.6}$ &
$    29.2_{-5.5}^{+6.6}$ &
257.0/239
\end{tabular}
\end{center}
\end{table}

The model parameters for the best-fitting parallax model are presented in Table 1.
The best fit has a $\chi^2=257$ for 239 degrees of freedom.
We found that the lens trajectory parameters  ($\u0$ and $\psi$) are not
well-specified in the ecliptic plane, very likely due to the fact
parallax signature is only modest for \sc33.
Fortunately, 
the most important lens parameters are well constrained, in particular,
we have
\beq
\rEt=29.2_{-5.5}^{+6.5} \,{\rm AU}, ~~
\tE=641_{-55}^{+70} \,{\rm day}, ~~
\Delta f=-240.9^{+1.4}_{-1.6}, ~~
\fl=43.3\pm 5.1.
\eeq
The Einstein radius crossing-time is about 641 days, the
largest ever reported for a microlensing event. The projected Einstein 
radius on the observer plane is also very large. As the flux in the
reference image is $f_R=359.5$ (Wo\'zniak et al. 2001), one sees that
the total baseline flux is therefore $f_0=f_R+\Delta f=118.6$
(cf. eq. \ref{eq:f}). The lensed star therefore only contributes
$\fl/f_0 \approx 36.5$ per cent of the total baseline flux. Note
the blending fraction is well constrained in the model. The
baseline {\it I}-band magnitude of the lensed star is 
about $18.1-2.5\log(\fl/f_0)=19.2\,{\rm mag}$.
The lensed star was highly magnified,
reaching a magnification of about $\Amax \approx 32$ at the peak.

The projected Einstein radius and the time-scale $\tE$ immediately allow us to derive
a transverse velocity projected onto the observer plane
\beq
\tilde{v} = \frac{\tilde{r}_{\rm E}}{\tE} = 79\pm 16 \mbox{ km s}^{-1}.
\eeq
The lens mass can be expressed as a function of the relative
lens-source distance (see Soszy\'nski et al. 2001; Gould 2000),
\beq \label{eq:limit}
M = \frac{c^2 \tilde{r}_{\rm E}^2}{4G} \left( {{1 \over \Dl}
- {1  \over \Ds} } \right)
=10.5 M_\odot \left( \frac{\tilde{r}_{\rm E}}{\rm29.2\, AU} \right)^2
\left( \frac{\pirel}{0.1 \rm mas} \right),
~~\pirel\equiv{{{\rm AU} \over \Dl} - {{\rm AU} \over \Ds} }.
\eeq
As can be seen from this equation, the lens mass depends on 
the relative lens-source
parallax, $\pirel$: If the source is about 7\,kpc away, and
the lens lies in the disc half-way between the observer and the 
source ($x=1/2$),
then $\pirel \approx 0.143$ mas, which gives a lens mass of about 
$15M_\odot$; as a comparison, for a bulge self-lensing event with 
$\Ds \approx 8\,\rm{kpc}$ and $\Dl \approx 6\, \rm{kpc}$,
then $\pirel \approx 0.042$ mas, which would give a lens mass of
about $4.4M_\odot$.
However, this latter scenario may be less likely since the projected 
velocity of the lens is relatively low (see \S4, Derue et al. 1999).
In either case, the implied lens mass seems to be rather large, well
beyond the measured mass for neutron stars ($1.4M_\odot$).

\section{Discussion}

We have systematically searched for parallax events in the 520 microlensing
candidates identified using the difference image analysis (Wo\'zniak et
al. 2001). In this process, we have discovered an extremely long
microlensing event with an Einstein radius crossing time
$\tE=641\,{\rm days}$,
the longest time-scale ever reported.
The event shows a small but significant parallax effect caused by the
motion of the Earth around the Sun. This
allows one to derive the projected Einstein radius on the observer plane
of $\rEt\approx 29.2$AU. We emphasize that while some parameters are not
well-constrained, the limit on $\rEt$ is quite robust, and it is important
to understand why. $\rEt$ is limited from below because the parallax
effect is quite subtle: a smaller $\rEt$ value would mean that
the Earth motion makes a larger relative excursion, and hence 
the perturbation on the light curve may become too large to be
compatible with observations. $\rEt$ is limited from above because
if it is too large, then the parallax model would become similar to the
standard model, i.e., it will be inconsistent with the data.
Somewhat paradoxically,
had the parallax effect been smaller than observed, the projected
Einstein radius on the observer plane would have to be even larger, 
implying an even larger lens mass.

In this paper, we have adopted the point source approximation, ignoring
the finite size of the lensed star. It is important to see whether the
assumption is justified particularly because the star was highly
magnified. The finite source size effect becomes important
when the closest approach is smaller than or comparable to the stellar radius
(Gould 1994; Witt \& Mao 1994; Nemiroff \& Wickramasinghe 1994). In the
source plane, the closest approach, $d$, is given by
\beq
d = \rEt {1-x \over x} {1 \over \Amax} \approx 200R_\odot  {1-x
\over x}, ~~ \Amax \gg 1
\eeq
 From the color-magnitude diagram (Fig. \ref{fig:cmd}),
the lensed star is likely to have
a stellar radius no more than a few solar radii. So the
closest approach is much larger than the stellar radius,
justifying the point source approximation.

The derived $\rEt$ and $\tE$ from the fitting allow us to express
the lens mass with a dependence on the relative lens-source parallax
(see eq. \ref{eq:limit}). If we assume the source is at $D_{\rm
s}=8\kpc$, then the lens mass only depends the parameter, $x$, 
the ratio of the distance to the lens and the distance to the source. 
The low projected velocity constrains the value of $x$. If the lens and
the Sun follow the pure galactic rotation, but the source is stationary
at the Galactic centre, then $\tilde{v_t}=220x/(1-x)\kms$.
The derived transverse velocity $\vt\approx 79\kms$
then implies $x \approx 0.26$, which in turn gives a lens mass of
$37.3M_\odot$. In principle, a maximum likelihood analysis on $x$
can be performed following Alcock et al. (1995) using the observed
velocity information.
However, such an analysis depends on the uncertain Galactic model (both
on the mass density distribution and the kinematics of stars). We do not
perform such a calculation here. We note, however, that
our lensed star is roughly in the same direction as theirs and has nearly
the same projected transverse velocity ($75\kms$),
so we expect to obtain a similar probability distribution for $x$;
their calculation indicates a value of x which is slightly smaller
than the na\"{\i}ve estimate above, and this would
imply a lens mass that is even larger. If a star with
$M > {\rm few}\,M_\odot$ is
still burning nuclear fuels, it will be much more luminous than
$I=18.1$. Hence, if the lens is indeed this massive,
then it must be dark, and it follows that
it is likely to be a stellar mass black hole.

\begin{figure}
\plotone{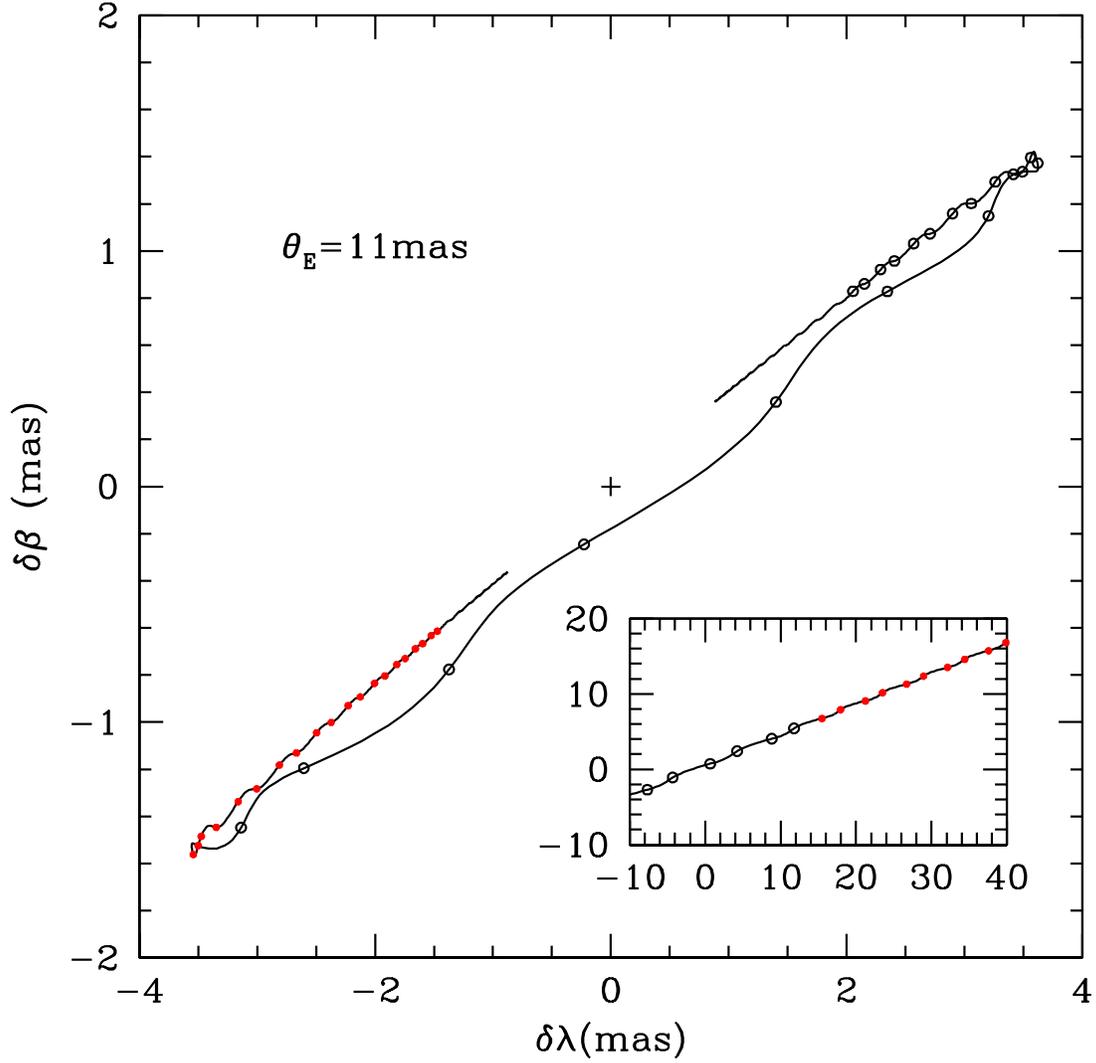}
\caption{The predicted astrometric motion of the light centroid
in the ecliptic plane. The size of the motion is proportional
to the angular Einstein radius, which we have 
taken to be $11\mas$ (see eq. \ref{eq:thetaE}).
The solid dots indicate the centroid positions
every six months on and after July 31, 2001 while
the open dots indicate the positions every six months before
the date. The plus sign indicates the source position when unlensed.
Notice that the scales on the two axes are different. The inset
shows the lens position relative to the centroid of the source in mas.
The open and filled circles have the same meaning as in the main panel.
\label{fig:astrometry}
}
\end{figure}

There may be a better and empirical method to test the
black hole hypothesis. While the photometric microlensing event is now over,
the astrometric microlensing signature is still ongoing, owing to the
much slower decay of the astrometric signature as a function of 
the impact parameter
(e.g. Gould 1992; Hosokawa et al. 1993;
H${\rm \o}$g, Novikov \& Polnarev 1995; Miyamoto \& Yoshi 1995; Walker 
1995; Paczy\'nski 1998). Ignoring the Earth's motion, 
the astrometric signature 
follows an ellipse. The major axis and minor axis are both
proportional to the angular Einstein radius, given by
\beq \label{eq:thetaE}
\thetaE = {\rEt \over \Ds} {1-x \over x}
\approx 3.7\, {\rm mas}\, {1-x \over x}\,
{8\kpc \over \Ds}.
\eeq
The predicted astrometric motion is not very well specified due
to the uncertainty in the trajectory. Fig.  \ref{fig:astrometry}
illustrates the prediction for the best-fitting model
with $x=0.25$ ($\thetaE=11\mas$). The origin of the astrometry
is chosen to be the position of the star when the lens is at infinity.
One sees that the astrometric motion is no longer
an ellipse due to the parallax effect. The largest
astrometric motion from the origin
is $\thetaE/\sqrt{8}\approx 3.9\mas$ for this case.
The soon-to-be installed Advance Camera
for Surveys \footnote{http://www.stsci.edu/cgi-bin/acs} on board HST
will be an ideal instrument for detecting this signature. The point
spread function is well sampled for this instrument, and 
it may be able to reach an astrometric accuracy as high as $0.1\mas$.
HST has another distinctive advantage over the ground based
interferometers as it can resolve the blends much more
easily. Multi-colour data from HST will also be useful
for studying the colour of the lensed star as currently only the {\it I}-band
photometry is available. However, the astrometric motion is quite
gradual and may be confused with the star's proper motion,
hence a multi-year monitoring project would have to be undertaken. 
Spectroscopic observations of the lensed source will be useful to
put further constraints on the lensing kinematics involved.
The lens may also be accreting interstellar gas, and could be
luminous in the X-ray. It would be very interesting to obtain a deep
image using sensitive X-ray satellites such as XMM-Newton and Chandra.
It will also be interesting to see whether the source is luminous in the
radio. Radio observations have distinct advantages, as it is not
affected by dust, and VLBI observations could reach $\sim \mas$
astrometry. The inset in Fig. \ref{fig:astrometry} shows the
position of the lens relative to the source centroid, which already reaches
tens of $\mas$ for our example. Such a shift, if detected, will be
a dramatic confirmation of our model.

\section*{Acknowledgement}

We acknowledge Bohdan Paczy\'nski for many stimulating
discussions and comments on the manuscript. 
SM was supported by the NSF grant AST-9820314 
as a visiting scientist to Princeton.
MCS acknowledges the financial support of a PPARC studentship.
PW was supported by the Laboratory Directed Research \& Development
funds (X1EM and XARF programs at LANL). 
This work was also supported by the polish grant KBN 2P03D01418.

{}


\end{document}